\documentstyle[11pt,aasms4]{article}

\def\alwaysmath#1{\ifmmode{#1}\else{$#1$}\fi}
\def\msun{\alwaysmath{{M}_{\odot}}}

\def\arcsec{\hbox{$^{\prime\prime}$}}
\def\etal{{et al.~}}

\def\ers{\alwaysmath{{\rm \, erg\,sec^{-1}}}}
\newcommand\hst{{\sl HST}}
\newcommand\HST{{\sl HST}}
\newcommand\rosat{{\sl ROSAT}}

\righthead{FERRARO, ET AL.}
\lefthead{UV OBSERVATIONS OF GLOBULAR CLUSTER M92}

\slugcomment{In press in the Astrophysical Journal}

\begin{document}

\title{Another Faint UV Object Associated with a Globular Cluster
X-Ray Source: The Case of M92\footnote{Based on
observations with the NASA/ESA {\it Hubble Space Telescope}, obtained at
the Space Telescope Science Institute, which is operated by AURA, Inc.,
under NASA contract NAS5-26555}}

\author{Francesco R. Ferraro\altaffilmark{2,3},
	Barbara Paltrinieri\altaffilmark{4,5},
	Robert T. Rood\altaffilmark{6},
	Flavio Fusi Pecci\altaffilmark{7,3},
        and
        Roberto Buonanno\altaffilmark{5}}

\altaffiltext{2}{
European Southern Observatory, Karl Schwarzschild Strasse 2,
D-85748 Garching bei M\"unchen, Germany; fferraro@eso.org}
\altaffiltext{3}{on leave from 
Osservatorio Astronomico di Bologna, via Ranzani 1, I-40126
Bologna, Italy;
ferraro@astbo3.bo.astro.it,
flavio@astbo3.bo.astro.it.}
 \altaffiltext{4}{Istituto di Astronomia -- Universit\'a La Sapienza,
via G.M. Lancisi 29, I-00161 Roma, Italy; barbara@coma.mporzio.astro.it}

\altaffiltext{5}{Osservatorio Astronomico di Roma, 00040 Monte Porzio, ITALY}

\altaffiltext{6}{Department of Astronomy, University of Virginia, 
P.O. Box 3818, Charlottesville, VA 22903-0818;
rtr@virginia.edu.}
 
\altaffiltext{6}{Stazione Astronomica di Cagliari, 09012 Capoterra, Italy.}

\begin{abstract}
The core of the metal poor Galactic Globular Cluster M92
(NGC 6341) has been observed with WFPC2 on the {\it Hubble Space
Telescope} through visual, blue and mid-UV filters in a program
devoted to study the evolved stellar population in a selected
sample of Galactic Globular Clusters.
In the UV $(m_{255}, m_{255}-U)$ color magnitude diagram
we have discovered a faint `UV-dominant' object. This star lies
within the error box  
of  a Low Luminosity Globular Cluster X-ray source (LLGCX)
recently found in the core of M92.
The properties of the
UV star discovered in M92 are very similar to those
of other UV stars found in the core of some
 clusters (M13, 47 Tuc, M80, etc)---
 all of them are brighter in the UV than in the visible and  are located
in the vicinity of a LLGCX.
We suggest that these stars are a new sub-class of cataclysmic variables.
\end{abstract}

\keywords{
stars: evolution ---
novae, cataclysmic variables ---
globular clusters: individual (M92) ---
ultraviolet: stars ---
}

\section{Introduction} \label{sec:intro}

Despite their rarity, it is the exotic creatures that attract the
crowds at the zoo; similarly the exotic objects in the stellar zoo
attract our attention. Unusual environments often lead to relatively
large populations of the exotic. So it is with the cores of the
Galactic Globular Cluster (GGCs), which have long been thought to
harbor a variety of exotic objects---blue stragglers, low mass X-ray
binaries, cataclysmic variables, millisecond pulsars, etc.  Most of
these objects are thought to result from   various kinds of binary
systems whose nature and even existence can be strongly affected by
dynamics in the dense cluster cores. 

When a binary system contains a compact object (like a neutron star or
white dwarf) and a close enough secondary, mass transfer can take
place. The streaming gas, its impact on the compact object, or the
presense of an accretion disk can give such systems observational
signatures which make them stand out above ordinary cluster
stars. These signatures might include X-ray emission, significant
radiation in the ultraviolet (UV), emission lines, or rapid time
variations. The first evidence for such objects in globular clusters
was the discovery of X-ray sources. One population of X-ray sources
with $ L_{X} > 10^{34.5} \ers $ (the so-called Low Mass X-ray
Binaries, LMXB) are thought to be binary systems with an accreting
neutron star because of their X-ray bursts.  LMXBs are very
overabundant (a factor 100) in GGCs with respect to the field,
presumably because the high stellar density has led to many capture
binaries.

Given the existence of neutron star systems in GGCs one might expect
to find many more analogous systems involving white dwarfs (WDs). In
the field, binary systems in which a WD is accreting material from a
late type dwarf, i.e., a main sequence or subgiant star, are observed
cataclysmic variables (CVs).  CVs are well-studied objects in the
field, where they are thought to form by the evolution of primordial
binaries. They come in many varieties depending on stellar masses,
mass transfer rates, magnetic field strength, etc.  In GGCs one can
expect even more variety because CVs located in dense clusters could
have been created by dynamical processes (Hut \& Verbunt, 1983, Bailyn
1995), while the CVs in low-density clusters result from primordial
binary systems (Verbunt \& Meylan 1988).

Numerical simulations (e.g., DiStefano \& Rappaport 1994) suggest that
$> 100$ white dwarf binaries might be found in massive clusters like
47~Tuc and $\omega$~Cen, and several 10's in more typical clusters.
Despite the expectation of large numbers of CV like stars, searches
have turned up only a relatively small number. Of course, part of the
problem arises because of the difficulty of search for rather faint
objects in crowded globular cluster fields.  However, exploiting the high
resolution of {\it HST}, it has become possible to search
GGC centers for several of the anticipated CV signatures. Still the
number of candidates is small:

\begin{itemize}

\item 
More than 30 low luminosity X-ray sources with $L_x < 10^{34.5} \ers$
(hereafter LLGCXs) have been discovered in 19 GGCs (Johnston \&
Verbunt 1994). Despite their relatively large numbers there is no
consensus model for LLGCXs (see Verbunt et al. 1994) and Hasinger,
Johnston \& Verbunt 1996). The fainter LLGCXs ($L_x < 10^{32}
\ers$) might well be associated with CVs (van Paradijs (1983), Hertz \&
Grindlay, 1983).

\item There are three objects connected with conventionally detected CVs:
a dwarf nova in M5 (Margon, Downes \& Gunn 1981); HST UV detections of
optical counterparts to a dwarf nova in 47 Tuc (Paresce \&
DeMarchi 1994) and possibly the
historical nova in M80 (Shara \& Drissen 1995)

 \item Using \HST, H$\alpha$ emission has been observed from three
objects in NGC 6397 (Cool et al. 1995), and two objects in NGC 6752
(Bailyn et al. 1996).

 \item Also using \HST, a number of candidate CVs have been
selected on the basis of UV excess and variability: the Einstein
dim source in 47 Tuc (Paresce, De Marchi \& Ferraro 1992) and few CVs
candidates in NGC 6624 (Sosin \& Cool 1995).

\end{itemize}

We are involved in two long-term HST projects to study in detail the
evolved populations in a sample of GGCs, at different wavelengths
ranging from the UV to the near IR.  Although we were not specifically
hunting for CVs, we have used the UV exposures to search for exotic
objects in the core of GGCs. This search has been very fruitful: in
our data-base (9 clusters) we have found in {\it all GGCs properly
observed (with exposures deep enough and in the right UV bands) there
is at least one faint object with a strong UV excess with respect to
the main stellar population of the cluster)}. These stars are brighter
in the UV than in the visible, so we will refer to them as
UV-dominant (UVD). We wish to carefully distinguish between these objects
and objects which are called `UV-excess' objects on the basis of their
colors in the visible or perhaps near UV. Previously we have reported
on three UVD stars in the GGC M13 (Ferraro et al. 1997).
Two of these objects have been found to lie within the error boxes of
LLGCXs (Fox et al. 1996), and we argue that they are excellent CV
candidates.  Here we report on the discovery of another faint
UVD star in the core of M92, and we suggest that this star is
physically connected to the X-ray emission detected in the cluster.

\section{Observations}

 \hst-WFPC2 frames were obtained on December 1995
 (Cycle 5 : GO 5969, PI: F.
Fusi Pecci).  We report here results
obtained using the deep exposures (3600 sec and 2200 sec)
through the U (F336W) and mid-UV (F255W) filters, respectively.
The CMDs presented here are  results of
the four WFPC2 chips (namely PC1, WF2, WF3, WF4), obtained with the PC
located on the cluster center (some further discussion of these
results can be found in Ferraro et al. 1997, 1998).

\begin{figure*}[htb]
\centerline{\null}
\vskip4truein
\includegraphics{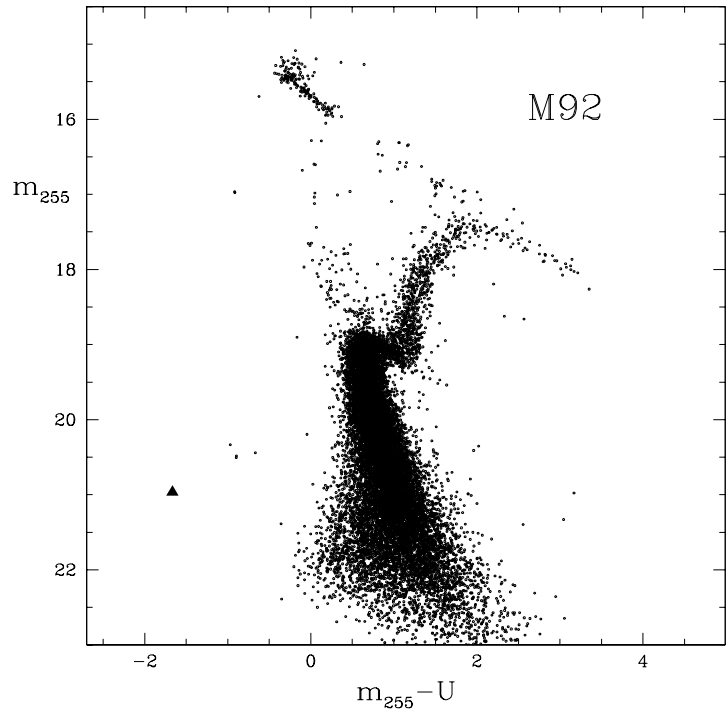}
\figcaption[fig1.ps]{
The $(m_{255},~m_{255}-U)$ color magnitude diagram for M92.
All stars detected in the WFPC2 field of view have been plotted.
The faint UV star is plotted as a large filled triangle.
\label{fig:cmd}}
\end{figure*}

All the reductions have been carried out using ROMAFOT (Buonanno et al
1983), a package specifically developed to perform accurate photometry
in crowded fields.  In order to identify the objects in each field we
used the median frame, obtained by combining all single exposures
in each color. The PSF-fitting procedure was then performed on each
individual frame separately and the instrumental magnitudes were then
averaged.   The instrumental magnitudes
have been converted to  fixed aperture photometry and where
appropriate calibrated to the Johnson system using equation 8 and
Table 7 in Holtzmann \etal\ (1995). F255W magnitudes have
been calibrated to the STMAG system using table 9 by Holtzmann et al.
(1995).

In this paper we adopt for M92 a distance modulus $(m-M)_0=14.78$
from Ferraro et al. (1999a) who have recently determined moduli for
 a sample of 61 GGCs within the framework of an homogeneous re-analysis
of the evolved sequences of the CMD.

\section{Results}

Figure 1 shows the $(m_{255},m_{255}-U)$-CMD for the global sample of
 stars detected in all the four WFPC2 chips. More than 20,000 stars
 have been measured in these filters in the global WFPC2 Field of
 View.  Inspection of this diagram shows that a few (5) blue low
 luminosity objects lie significantly outside the main loci defined
 by the majority of the cluster stars.  Four of them
 are clumped at $(m_{255}-U)\sim-1$, and one object (namely star
 $\# 8203$) shows a very strong UV color $(m_{255}-U)<-1.5$. This object
 is at only $15\farcs 7$ from the cluster center (assumed at $\alpha
 _{2000}=17^{\rm h}\, 17^{\rm m}\, 07\fs 3 ,~\delta_{2000}=43\arcdeg\,
 08\arcmin\, 11\farcs0$, Djorgovski \& Meylan 1993), just outside
 the core radius of the cluster ($r_c=14\arcsec$).  It is located
 at the extreme Northern edge of the PC chip.  This region is still
 very crowded, and the UV star is close (less than $1\arcsec$) to an
 HB star and a very bright giant ($V\sim 12.8$) which is heavily
 saturated in the $V$ and $I$ deep exposures. Hence, only measures in
 the $F255W$ and $U$ filters are possible.

\begin{table*}[htb]
\caption{\hfil UV star and X-ray source in the core of M92
 \label{tab:srcs} \hfil}
\begin{center}
\begin{tabular}{lllccc}
\tableline
\tableline
  &  $\alpha_{2000}$   & $\delta_{2000}$  & $U$ & $m_{255}$ &  $L_x$\\
\tableline
8203 &$17^{\rm h}\, 17^{\rm m}\, 06\fs 559$ 
& $43\arcdeg\, 08\arcmin\, 24\farcs 397$& 22.63& 20.97& --\\
M92X-C& $17^{\rm h}\, 17^{\rm m}\, 06\fs 3$ & 
$43\arcdeg\, 08\arcmin\,
23\farcs0$ &  --  & -- & $4.6 \times 10^{32}$\\ 
\tableline
\end{tabular}
\end{center}
\end{table*}
 
  The
possible variability of the  UV source was examined by analyzing
each available frame 
separately. No clear indication of variability was revealed
from this analysis, but we cannot strongly exclude this possibility
since our observations do not have much time coverage.

\begin{figure*}[htb]
\vskip4.5truein
\includegraphics{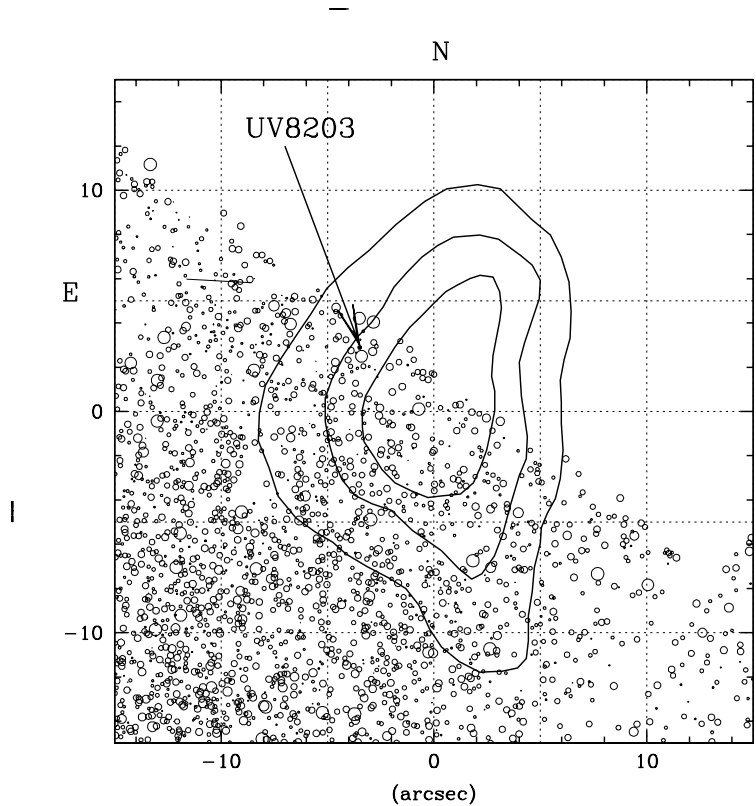}
\caption[fig2.ps]{
Map of the image taken through F255W filter in the
region of the X-ray source M92C [located at (0, 0)].
The $x$ and $y$
scales are in arcsec. The X-ray emission contour map has been overplotted.
The location of the  UV-star discovered in this study is indicated by
an arrow.
 \label{fig:map}}
\end{figure*}

Fox \etal\ (1996) recently presented \rosat\ High Resolution Imager
(HRI) observations of M92 and identified 7 low luminosity X-ray
sources in the field of view of the cluster.  In particular, they drew
attention to the X-ray source found in the core, M92C (hereafter
M92X-C). M92X-C is only $\sim 17\arcsec$ from the cluster center and
has a high probability ($\sim 99.8\%$) of being associated with the
cluster.  The LLGCX is also located at the Northern edge of the PC in our
HST field of view just in the region where the UV star has been
detected.  Figure 2 shows a region of $\sim 30\arcsec \times
30\arcsec$ centered on the nominal position of the M92X-C.  The
contours of the X-ray emission (from Figure 2 by Fox \etal) have been
overplotted on a digital map of the $F255W$ image.  The absolute
positions, the observed magnitude and the X-ray flux for the UV star
and the X-ray source are listed in Table 1. Note that the X-ray
luminosity has been properly scaled in order to take into account the
different distance modulus adopted here with respect to that used by
Fox et al. (1996).

Because the UV object is $\la 4.5\arcsec$ from   the nominal position
of the X-ray emission, we strongly suggest a physical connection
between the two.  Note that the 4 UV objects located in the CMD at
$(m_{255}-U)\sim-1$ are much more distant ($d>37\arcsec$) from the
X-ray source.
  
\section{Discussion}

\begin{table*}[htb]
\caption{\hfil UV stars associated with LLGCXs
 \label{tab:uvstars} \hfil}
\footnotesize
\begin{center}
\begin{tabular}{llcccccccccccc}
\tableline
\tableline
 Cluster &  Name & $(m-M)_0$ & $E(B-V)$ &$M_I$ & 
$M_V $ & $M_B$ & $M_U$ & $M_{255}$ & $M_{160}$ &
$M_{140}$& $L_x(10^{32})$ & range in KeV\\
\tableline
M92 & 8203 & 14.78 & 0.02 &-- &  -- & -- & 7.75& 6.05 & -- & --
&4.6&0.1--2.4\\
M13 & 23081 & 14.43 & 0.02& --&  7.69 & -- & 6.09 & -- & 2.36 & --&
7.4&0.1--2.4\\
M13 & 21429 & 14.43 & 0.02 &-- & 6.55 & -- & 5.33 & 3.92& 2.72 & --
& 2.9&0.1--2.4\\
M80 &       &  14.96 & 0.18&-- &  --  & 6.52 & 5.47 & -- & -- & --&
7.6&0.5--2.5\\
M5   &    V101 & 14.37 & 0.03 &  -- &5.80 & 6.2&3.7& --& --& --& 1.1&0.5--2.5\\
NGC6397 & CV1 & 11.92 &0.18& 6.44 &5.75 & 6.31 & 5.47 & -- & -- & --&
1.1&0.1--2.0\\
NGC6397 & CV2 & 11.92 & 0.18& 7.70 & 6.93 & 7.65 & 6.38 & -- & -- & --&
0.9&0.1--2.0\\
NGC6397 & CV3 & 11.92 & 0.18& 8.42 & 7.64 & 7.80 & 6.29 & -- & -- & --&
 0.8&0.1--2.0\\
47 Tuc & V1 & 13.32 & 0.04& -- & --&--&  6.08 & 4.93 & 4.36 & --& 8.0
&0.5--2.5 \\
47 Tuc & V2 & 13.32 & 0.04& --  &--& 7.52 & 6.58 & 6.03 & -- & -- &
1.3 &0.5--2.5\\
NGC6752 & 1 & 13.18 &0.04&6.05 &  7.25 & 7.66 & --& 7.23& --& --& 1.7 & 0.1-2.4\\
NGC6752 & 2 & 
13.18 &0.04&  6.46  &7.59&8.16 & --&-- & --& --& 1.7 & 0.1-2.4\\
\tableline
\end{tabular}
\end{center}
\end{table*}

Faint UV stars (similar to UV8203) have been discovered in the core of
other GGCs, and some of them have been found to be nearly coincident
with X-ray sources.  The discovery reported in this paper makes the
possibility of a chance coincidence of UV objects with the X-ray
sources appear even less likely.     
With the strengthening evidence for a (physical) connection between
UV objects and LLGCXs, it is appropriate to
compare the photometric properties of a sample of faint UV stars
found in the vicinity of LLGCX in GGCs. In doing this
we select some of the most   
recent findings:

\begin{description}

\item{{\bf M13}---} Ferraro et al. (1997) using deep UV-HST
observations found three extremely blue, low luminosity
objects in the very central region of M13.  Two of them are nearly
coincident with the two-peaked X-ray emission detected by Fox et al
(1996).

\item{{\bf M5}---} V101 in M5 was discovered by Oosterhoff (1941), who first
suggested that it could be a CV (dwarf nova). Spectroscopic 
(Margon, Downes \& Gunn 1981 and Naylor et al. 1989) and 
photometric  observations (Shara, Potter \& Moffat 1987) confirmed
this suggestion. Hakala et al. (1997) detected X-ray emission
associated with this object using using the ROSAT-HRI.

\item{{\bf M80}---} Shara \& Drissen (1995)
have  found two UV
objects in the globular cluster M80,
one of them might be associated with T~Sco, a
nova observed in 1860.  
This object might be connected with a 
 LLGCX located at  $8\arcsec$ from the UV object, although,
as suggested by Hakala \etal\ (1997),
 the position from the \rosat\ PSPC is not accurate
enough for a definitive identification.

\item{{\bf NGC 6397}---} Cool et al. (1998, C98) found a population of
7 UV stars in the core of NGC6397. They divided the UV star sample in
two subgroups: 4 stars showing variability and UV excess (in the sense
that they appear to be blue in the $U$ band but are indistinguishable
from MS stars in the $V,~V-I$ plane), and three non-variable UV
excess stars which are significantly hotter than the main-sequence in
all observed CMD planes. They call this second class `nonflickerers'
(NF).  The four variable stars are all within the ROSAT HRI X-ray
error circle and have been confirmed to be CVs (Cool et al. 1995,
Edmonds et al. 1999). However, they noted that {\it ``...two out three
NFs are outside the error circles of the three central X-ray sources
detected with ROSAT by Cool et al. (1993).''}  While C98 claimed that
the NF stars were a {\it new} class of faint UV stars, it is worth
noting that they are very similar to the three UV stars found one year
before in M13 by Ferraro et al. (1997).

\item{{\bf 47 Tuc}---} At least two objects with a strong UV excess
have been identified in the error box of X-ray source in the center of
47 Tuc: V1 (Paresce, DeMarchi \& Ferraro 1992) and V2 (a blue variable
discovered by Paresce \& DeMarchi 1994 (see also Shara et al. 1996).
V1 lies within the error circle of the X-ray source $X0021.8-7221$
detected by Einstein (Bailyn et al. 1988) and in the vicinity of a low
luminosity X-ray source (X9 in Table 2 of Verbunt \& Hasinger 1998).
Verbunt \& Hasinger (1998) also identified V2 as a candidate optical
counterpart of their source X19.

\item{{\bf NGC~6752}---} Bailyn et al. (1996) report the identification
of two candidate CVs in NGC~6752. Both stars fall at the edge of the
error circle of the X-ray source identified as B by Grindlay \& Cool
(1996) (flux given in Grindlay 1993). These stars are plotted as empty
circles in Figure 3.  Contrary to the behavior shown by all the other
objects candidates they have a quite strong rise up toward red
wavelengths suggesting that these objects have quite different spectral
characteristics with respect to the other objects listed above.  
Their position in the CMD (see Figure 3 by Bailyn et al) resemble
the CVs found by C98 in NGC~6397, however  observations  at wavelength
shorter than $B$ are needed in order to better constrain the
spectral behavior of these stars. For these reasons we
exclude them from the following discussion.

\end{description}

The absolute magnitude in different photometric bands for the 10 UV
stars possibly connected with X-ray emission in the 8 GGCs quoted
above are listed in Table 2.  Also reported are the adopted distance
moduli and reddening from Ferraro et al. (1999b).  Note that the
distance moduli adopted typically differ from earlier papers, for
example, for NGC~6397 is $\Delta(m-M)_V \sim 0.2$ mag larger than that
adopted by Cool et al. (1998). All the absolute magnitudes and X-ray
luminosities in Table~1 have been corrected accordingly. Note also
that the X-ray fluxes were determined over different energy ranges
which are given in the last column.

\begin{figure*}[htb]
\vskip3.8truein
\includegraphics{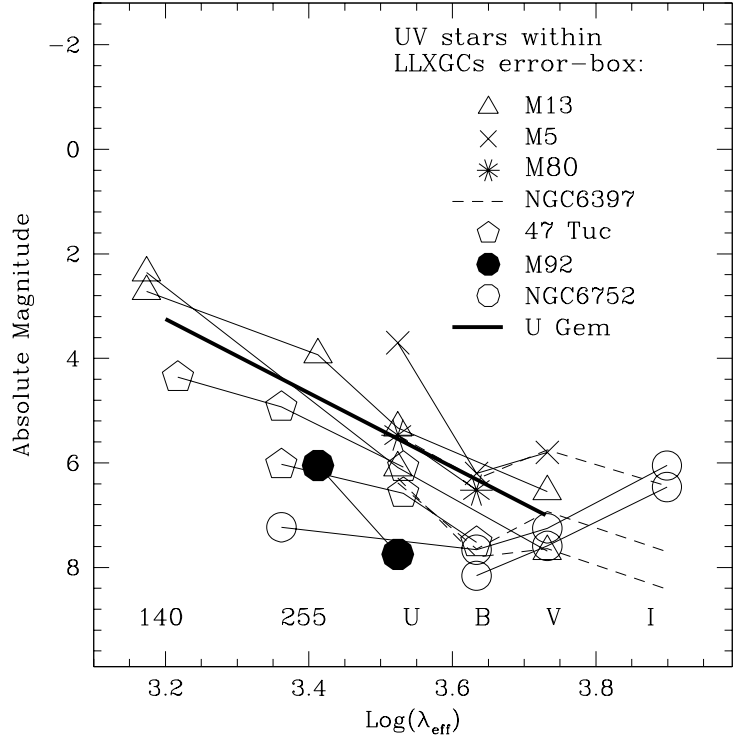}
\caption[fig3.ps]{
Comparison between the photometric characteristics of the UV stars discovered
in M92 and a sample of UV stars found within the error boxes of LLGCXs in
other GGCs (see Table 2). The absolute magnitudes obtained through
 various filters are plotted as function of the filter effective wavelength.
  Different symbols refer  to different clusters. The approximate location  of each filter
in terms of effective 
wavelength is labeled at the bottom of the figure.
 \label{fig:map}}
\end{figure*}

\begin{figure*}[htb]
\vskip3.7truein
\includegraphics{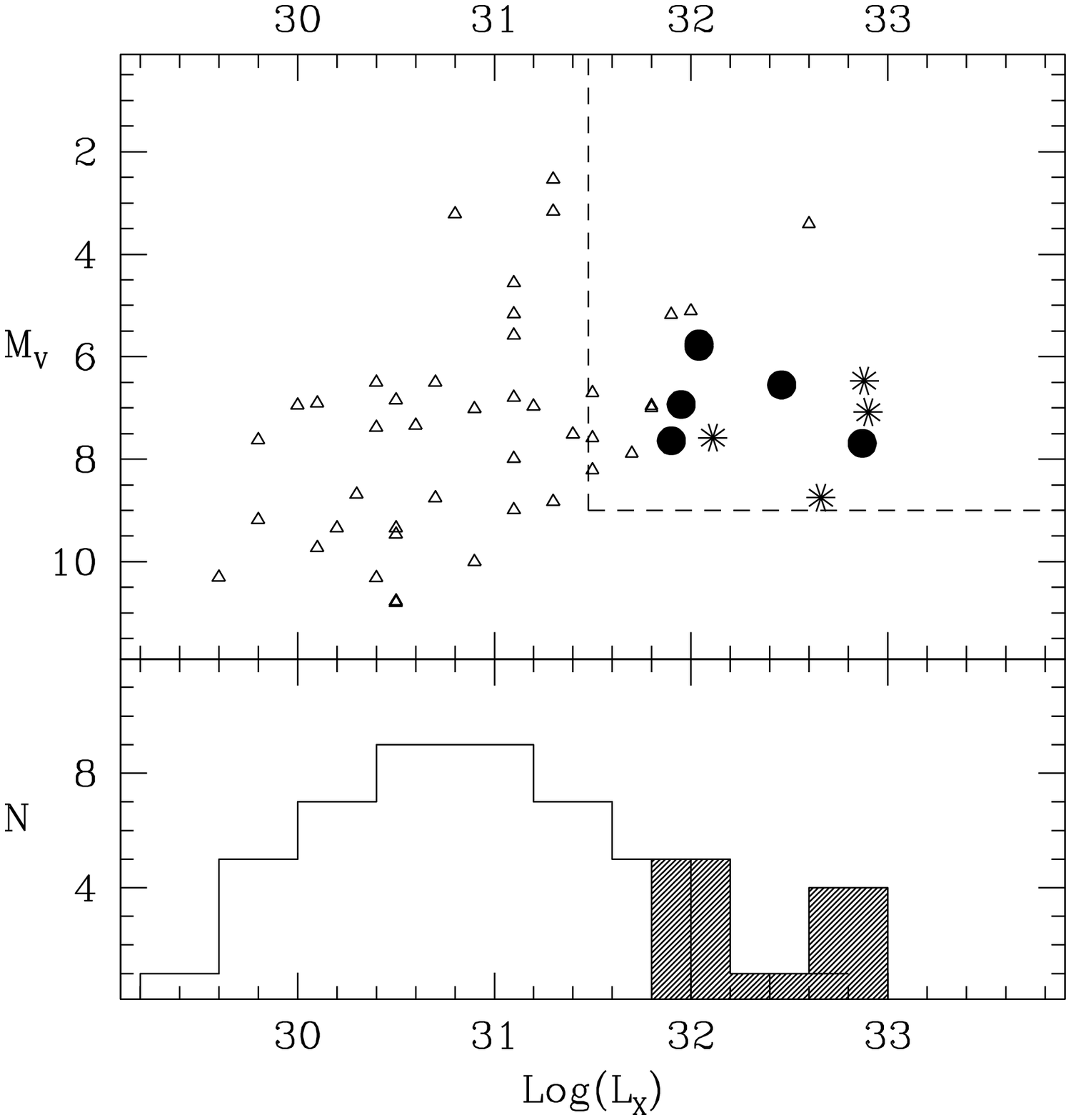}
\caption[fig4.ps]{The  properties
for  field CVs are compared with the candidates found in GGCs.
The upper panel shows  the absolute $V$ magnitude ($M_V$) 
as function of the X-ray luminosity ($\log\,L_x$).   
Field CVs  from Table 1 by V97, are plotted  as small open triangles, and
the UVD stars in GGCs listed in Table 2 are plotted as
large filled circles.  The large asterisks are the 4 candidates in Table 2 
for which no $V$ magnitude was available: it has been computed 
from $M_U$ assuming a mean color  $\langle U-V\rangle =-1$ (see text),
Dashed lines show approximate seaech limits for UVD stars/LLGCXs
in M13, M92 and NGC6397.
The lower panel shows the X-ray luminosity distribution
for  objects found in GGCs (shaded histogram) compared with the distribution
for the field CVs. 
\label{fig:map}}
\end{figure*}

In Figure 3 we plot the absolute magnitude for each star listed in
Table 2 as a function of the filter's effective wavelength. In the
figure different symbols refer to different clusters: open triangles
for M13; filled circles for M92, asterisks for M80, empty pentagons
for 47 Tuc, large $X$ for M5 and dashed lines for the three CVs in NGC
6397. For comparison the energy distribution for the field CV U~Gem is
plotted as an heavy solid line. Figure 3 provides an easy way to make
quantitative comparison.  In particular, all of the UV selected stars
show the same overall spectral trend, and we conclude that some of the
photometric properties of faint UVD stars associated with LLGCXs are
in reasonable agreement with each other {\it and} U~Gem.  It is
interesting to note that the slope of UV8203 in M92 and V101 in M5
appear to be steeper than the others, however far-UV observations are
required in order to confirm this impression.

 Though small, the sample listed in Table 2 can still be used to
derive some average properties of these objects, for example the
absolute $V,~B,~U$ magnitude, and $U-V$, $U-B$ colors which are used
to characterize CVs in the field.  From the data in Table 2 these
figures turn to be: $\langle M_V\rangle=6.7\pm0.8$, $\langle
M_B\rangle=7.0\pm0.7$, $\langle M_U\rangle=5.9\pm1.0$, $\langle
U-B\rangle=-1.1\pm0.3$ and $\langle U-V\rangle=-1.0\pm0.5$,
respectively.  These figures are in good agreement with the typical
absolute magnitude and colors for CVs in the field ($\langle
M_V\rangle \sim +7$, for the dwarf novae CVs; see van Paradijs 1983).

We may push our working hypothesis that the UVD objects are
physically associated with the X-ray sources further.  The X-ray
luminosity of each source is listed in the penultimate column of Table
2. As can be seen they have comparable X-ray luminosity, within the
range 1--$8 \times 10^{32}\,{\rm erg\,sec^{-1}}$ (with a mean value of
$\langle L_x\rangle=4\pm3 \times 10^{32}\,{\rm erg\,sec^{-1}}$).  
 
We can compare the observed photometric characteristics of the UV
objects found in GGCs with those obtained for field CVs using the
values listed in Table 2 along with data from Table 1 of the recent
compilation by Verbunt et al. 1997 (hereafter V97) who presented a
catalog with 91 CVs in the field detected during the ROSAT All Sky
Survey.  The upper panel of Figure~4 shows the absolute $V$ magnitude
as a function of the X-ray luminosity for all field CVs with known
distances listed by V97. These are plotted as small empty triangles.
The 6 UVD stars found in GGCs for which the $V$ magnitude has
been measured are plotted as large filled circles.  The $V$ magnitude
for the 4 stars for which no $V$ magnitude was directly measured has
been computed from $M_U$ assuming a mean color $\langle U-V\rangle=-1$
(see above). These are plotted as large asterisks.  The lower panel
shows the X-ray luminosity distribution for the GGC objects (shaded
histogram) compared with the distribution for the field CVs.  This
figure clearly shows that while the absolute $V$ magnitude for the
candidates in clusters are fully consistent with the field CVs, the
X-ray emission for CVs in GGCs seems systematically higher (as already
suggested by Ferraro et al. 1997, see also Figure 4 in Verbunt \&
Hasinger 1998) indicating that the X-ray luminosity of objects found
in GGCs is high relative to the visible compared to similar objects in
the field.

There is a strong observational selection effect at work in
Figure~4. The depth reached by the X-ray observations is typically
only a factor of two or three below the level of the detected
sources. The deepest is for NGC~6397 (Cool \etal\ 1993) which reaches
$L_x = 3 \times 10^{31}\,{\rm erg\,sec^{-1}}$. Dashed lines in Figure
4 show the maximum depth reached in surveys for UVD stars/LLGCXs in M13,
M92, \& NGC~6397. Figure~8 of V97 shows that the vast majority of
field CVs of all types are less luminous than this. A comparison of
the $M_V$ for the optically identified LLGCXs with typical values from
Warner (1987) shows that current optical surveys would also have
missed many field CVs if they were located in GGCs. The relatively
high $L_x$ and $L_{\rm opt}$ for LLGCXs could arise solely because
that is all that surveys to date could detect.

This bias has important consequences. Hakala et al. (1997) used the
high $L_x/L_{\rm opt}$ value of T~Sco in M80 to argue that it was not
associated with the candidate object suggested by Shara \&
Drissen. However, we now see that $L_x/L_{\rm opt}$ is systematically
higher for GGC CVs as compared to their field sisters, and thus, the
Hakala \etal\ argument is not valid.
 
 Only the three field CVs with highest $L_x$ are consistent with the
 GGC CV candidates in Figure~4. These are the DQ~Her systems
 V1223~Sgr, AO~Psc and TV~Col. These are strongly magnetic CVs of a
 class referred to as intermediate polars (IP).  Grindlay (1999)
 suggests that IPs might dominate the ROSAT survey since they are
 expected to have a ratio $F_x/F_{\rm opt}$ greater than non-magnetic
 CVs.  The problem with connecting the UVD stars with magnetic
 CVs is that the magnetic field might truncate the inner portion of
 the accretion disk (Grindlay 1999).  Thus, MCVs might not be expected
 to have strong UV-excess.
        
It is worth noting that the V97 $L_x$ distributions for field dwarf
novae (SU~UMa, Z~Cam, U~Gem stars) are based on small samples (11, 7,
7) and still span one to two orders of magnitudes. Larger samples could
well reach into the range observed in GGCs. Indeed, since we could
well be observing the bright end of a sample of several tens of
assorted types of CVs in a GGC, the large observed values of $L_x$
probably do not provide a significant constraint on CV types.

Perhaps the most solid detection of CVs in a GGC are those in NGC~6397
(Edmonds \etal\ 1999; C98). There are several
indications of mass exchange: H$\alpha$ emission, X-ray radiation,
spectra with emission lines, and time variability or flickering of the
optical radiation. The colors of these objects become redder as the
wavelength of filters employed increases. This suggests in, for
example, $V$ and $I$ the light from the cool secondary dominates a
rather weak accretion disk. This is one of the factors which led
Edmonds \etal\ (1999) and Grindlay (1999) to associate these objects
with the intermediate polar class of magnetic CV. The CVs in NGC~6752
are probably similar objects.

The UVD objects are clearly not this sort of
creature. Many have properties one might expect from a generic CV,
i.e., white dwarf/main sequence binaries with some mass exchange. They
share many properties with some field CVs---X-ray luminosity, UV
colors, absolute visual magnitudes. The facts that they tend to lie at
the extremes of the distributions and that some do not have detected
X-radiation may simply be a consequence of the rather low sensitivity
of current surveys. In this scenario the difference between the UV
dominant CVs and those in NGC~6397 is simply a matter of the relative
importance of light from the secondary and the accretion disk. For the
GGC UVD stars, as with most CVs in the field, the accretion disk
significantly outshines the secondary star (see for example Figure
2.16--2.17 in Warner 1995).

What other options are there? It might be tempting to extend either
the horizontal branch (HB) or white dwarf (WD) sequences into the
region of the UVD stars. However, the HB terminates at the helium
burning main sequence. Particularly for M13 where the observed HB
extends almost to its termination, we see that the UVD stars are
significantly fainter (Figure~1 of Ferraro \etal\ 1997). The ordinary,
i.e., carbon/oxygen, WD sequence is well observed, for instance, in M~4
 and all the observed WDs are fainter than
$M_U \sim 8.5$ (see Figure 6 by Richer et al. 1997),
$\sim 1$ mag. fainter than the faintest UVD in Table 2. 
 So the
UVD stars are not HB stars or carbon/oxygen white dwarfs.
 
Edmonds \etal\ (1999) have shown that at least one UVD star is not a
CV. Their \hst-FOS spectrum of one of the NF-UV objects identified by
C98 in NGC~6397 suggests a very hot high gravity object. They argue
that this is a low mass ($\sim 0.25\msun$) helium white dwarf. Such an
object could arise from either mass exchange in a binary system shortly  
after the primary leaves the main sequence or via stellar
collisions. The object has a velocity of $\sim 250\,{\rm km\,s^{-1}}$
relative to the cluster CVs which is most easily explained if it still
is in a binary system with a dark companion. (It might be
worth noting that NGC~6397 is a post-core-collapse cluster with a very
dense core, whereas the UVD objects we have found are in the moderate
density clusters M13 \& M92.) Could He WDs account for all GGC UVD
stars? This is certainly not the case for the objects in M5, M80, \&
47~Tuc which have shown variability consistent with known CV types. He
WDs would not produce X-rays in significant quantities. The chance
association of three LLGCXs with such rare objects as the UVD objects
we have found in M13 \& M92 would seem unlikely.

 \section{Conclusions}

We have now identified three UV dominant objects which appear to be
associated with X-ray sources in the GGCs M13 and M92. We argue that
these are generically CVs, i.e., white dwarf/main sequence binaries
with some mass exchange. They share many properties with some field
CVs---X-ray luminosity, UV colors, absolute visual
magnitudes---although they tend to lie at the extremes of the
distributions. The relatively high optical, UV, \& X-ray luminosities
are consistent with the notion that current surveys do not reach deep
enough to detect most of the CVs in GGCs. On the other hand, cluster
CVs are  older and live in a dramatically different environment from
their sisters in the field. Thus, it seems reasonable that cluster
objects might be a new class of CVs with properties which slightly
differ from those in the field.

The CVs found in NGC~6397 and NGC~6752 differ significantly from our
objects. Some of these have little or no UV excess. This should not be
surprising. The searches in NGC~6397 and NGC~6752 relied on H$\alpha$
and $R$ band, whereas we used the UV. There is considerable variety in
the properties of field CVs and no reason to suspect less variety in
GGC CVs.  Different search techniques operating at the margin of
detectability will certainly turn up different kinds of objects.

The UVD objects in GGCs probably come in several varieties. (Our GGC
projects do not seem to come up with simple answers.) Edmonds
\etal\ (1999) have shown that one is a hot high gravity object,
arguablly a He-WD. 
 Still we suspect that
most of these will turn out to be CVs, and that many UVD objects with
no X-radiation detected to date will show up as LLGCXs in more
sensitive X-ray surveys. This suspicion is fueled by the belief that a
significant population of generic CVs must be present in GGCs.

We have yet to make an  observation directly showing the hot diffuse
gas which would be the definitive evidence that our UVD objects are
CVs. HST STIS spectra could give such evidence. It would be extremely
valuable to develop a technique to identify GGC CVs using UV
photometry. Cluster cores are very congested in the red and
H$\alpha$/$R$ band searches will obviously be incomplete because of
the interference by bright red giants (see Figure~2 of Bailyn \etal\
1996). On the other hand the core of even the densest clusters are
relatively open in the UV (see Figure~1 of Ferraro \etal 1999b)

\acknowledgments
This research was partially supported by the {\it
Agenzia Spaziale Italiana} (ASI) and by the MURST as part of the project
{\it  Dynamics and Stellar Evolution in Globular Clusters}.
 F. R. F. acknowledges the {\it ESO
Visiting Program} for the hospitality.  
  R. T. R. is supported in part by
NASA Long Term Space Astrophysics Grant NAG 5-6403 and STScI/NASA
Grant GO-6607.


\begin{references}

\reference{bai88}
Bailyn, C.D., Grindlay, J.E., Chon, H., \& Lugger, P.M. 1988, ApJ, 331, 303  

\reference{bai95}
Bailyn, C.D. 1995, ARA\&A, 33, 133

\reference{bai96}
Bailyn, C.D., Rubenstein, E.P., Slavin, S.D., Cohn, H., Lugger, P.,
Cool, A.M., \& Grindlay, J.E. 1996,  ApJ, 473, L31


\reference{buo83} 
Buonanno, R., Buscema, G., Corsi, C.E.,
Ferraro, I.,  \& Iannicola, G. 1983, A\&A, 126, 278  




\reference{cool93}
Cool, A.M., Grindlay, J.E., Krockernberger, M., \& Bailyn, C.D. 1993, 
ApJ, 410, L103

\reference{cool95}
Cool, A.M., Grindlay, J.E., Cohn, H.N., Lugger, P.J., \& Slavin, S.D.
1995, ApJ, 439, 695


\reference{cool98}
Cool, A.M., Grindlay, J.E., Cohn, H.N., Lugger, P.J., \& Bailyn, C.D. 1998, 
ApJ, 492, L75 (C98)

\reference{dm93} Djorgovski, S., Meylan, G., 
1993, in Structure and Dynamics of Globular Clusters, ed. 
S. G. Djorgovski \& G. Meylan (ASP: San Francisco), 325 

\reference{dr94} DiStefano, R., Rappaport, S., 1994, ApJ, 437,733

\reference{edmunds99} Edmonds, P. D., Grindlay, J. E., Cool, A., Cohn,
H., Lugger, P., \& Bailyn, C. 1999, \apj, 516, 250

\reference{f97}
Ferraro, F.R., Paltrinieri, B., Fusi Pecci, F., Dorman, B., \& Rood, R.T.
1997, MNRAS, 292, L45

\reference{f98}
Ferraro, F.R., Paltrinieri, B., Fusi Pecci, F., Rood, R.T., \& 
Dorman, B. 1998, in
{\it Ultraviolet Astrophysics--Beyond the IUE Final Archive},
eds. R.  Gonz\'alez-Riestra, W. Wamsteker, \& R. A.  Harris (ESA:
Noordwijk), 561

\reference{f99a}
Ferraro, F. R., Messineo, M., Fusi Pecci, F., 
De Palo, M.A., Straniero, O., Chieffi, A., \&
Limongi, M. 1999a, AJ, 118, 1738


\reference{f99b}
Ferraro, F.R., Paltrinieri, B.,  Rood, R.T., \& 
Dorman, B. 1999b, ApJ, 522, 983
 
\reference{fox}
Fox, D., Lewin, W., Margon, B., van Paradijs, J., \& Verbunt, F. 1996,
MNRAS, 282, 1027

\reference{grindlay93} Grindlay, J. E. 1993, in Structure and Dynamics of Globular Clusters, ed. 
S. G. Djorgovski \& G. Meylan (ASP: San Francisco), 285

\reference{grindlay99}Grindlay, J. E. 1999, in Annapolis Workshop on
Magnetic CVs, ed. C. Hellier \& K. Mukai (ASP: San Francisco), 377


\reference{ha97}
Hakala, P.J., Charles, P.A., Johnson, H.M., \& Verbunt, F. 1997, MNRAS, 285, 693

\reference{hjv94} 
Hasinger, G., Johnston, H.M., \& Verbunt, F. 1994, A\&A, 288, 466

\reference{hg83}
Hertz, P., \& Grindlay, J.E. 1983, ApJ, 267, L83

\reference{holt1995}
Holtzmann, J.A., Burrows, C.J., Casertano, S., Hester, J.J.,
 Trauger, J.T., Watson, A.M., \& Worthey, G., 1995, PASP, 107, 1065

\reference{hv83}
Hut, P., \& Verbunt, F. 1983, Nature, 301, 587

\reference{jv96} Johnston, H.M., \& Verbunt, F. 1996, A\&A, 312, 80.

\reference{mdg81}
Margon, B., Downes, R.A., \& Gunn, J.E. 1981, ApJ, 247, L89

\reference{nay89}
Naylor, T., Allington-Smith, J., Callanan, P.J., Hassall, B.J.M., Machin, G., 
Charles, P.A., Mason, K.O., Smale, A.P., \& van Paradijs, J. 1989, MNRAS, 241, 25

\reference{Oo41} Oosterhoff, P.T. 1941, Annual Sternw. Leiden, 174, 1

\reference{pdf92}
Paresce, F., De Marchi, G., \& Ferraro, F.R. 1992, Nature, 360, 46

\reference{pardem}
Paresce, F., \& De Marchi, G. 1994, ApJ, 427, L33

\reference{ric97} Richer, H.B., et al. 1997, AJ, 484, 741

\reference{spm87}
Shara, M.M., Potter, M., \& Moffat, A.F.J. 1987, AJ, 94, 357 

\reference{sd95}
Shara, M.M., \& Drissen, L. 1995, ApJ, 448, 203

\reference{sd96}
Shara, M.M., Bergeron, L.E., Gilliland, R.L., Saha, A., \& Petro, L. 1996, 
ApJ, 471, 804

\reference{sc95}
Sosin, C., \& Cool, A.M. 1995, ApJ, 452, L29


\reference{para83}
van Paradijs, J. 1983, in Lewin W. H. G., Van den
Heuvel E. P. J., eds, Accretion Driven Stellar X-ray Sources.
CUP, Cambridge,  p. 189

\reference{vm88}
Verbunt, F., \& Meylan, G. 1988, A\&A, 203, 297

\reference{ver94}
Verbunt, F., Johnston, H.M., Hasinger, G., Belloni, T., \& Bunk, W. 1994, in
Shafter A. W., ed, ASP Conf Ser. 56, Interacting Binary Stars. ASP, San
Francisco, p. 244

\reference{} Verbunt, F., Bunk, W.H., Ritter, H., \& Pfeffermann, E. 1997, 
A\&A, 327, 613 (V97).

\reference{} Verbunt, F., \& Hasinger, G. 1998, A\&A, 336, 895.

\reference{} Warner, B., 1987, MNRAS, 227, 23

\reference{} Warner, B., 1995, in Cataclysmic Variable Stars, Cambridge University Press, Cambridge

\end{references}
\end{document}